\documentclass[preprint,preprintnumbers,amsmath,amssymb]{revtex4}
\usepackage{graphicx}
\usepackage{dcolumn}
\usepackage{bm}

\begin{document}

\title{Quantum information processing in a spin-bus system of coupled chains}

\author{Xiang Hao}

\author{Shiqun Zhu}
\altaffiliation{Corresponding author} \email{szhu@suda.edu.cn}

\affiliation{School of Physical Science and Technology, Suzhou
University, Suzhou, Jiangsu 215006, People's Republic of China}

\begin{abstract}

The effective Heisenberg interaction of long distance is constructed
in spin qubits connected to a bus of two strongly coupled chains.
Universal quantum computation can be realized on the basis of the
bus which always keeps frozen at the ground state. It is found that
the effective interaction is primarily determined by the energy
spectra of the bus. With the variation of the distance between two
connecting nodes, the interaction alternately occurs between
antiferromagnetic and ferromagnetic ones. The long range interaction
can also be attained in coupled infinite chains. The quantum gate
operations with the high precision are implemented in the condition
of quantum fluctuations.

\end{abstract}

\maketitle

\section{Introduction}

The proposals for quantum information processing have important
developments in the spin qubits of quantum dots
\cite{Divin98,Kane98}. As the heart of solid state quantum
computation, the forms of the exchange interaction between qubits
have been recognized from the isotropic Heisenberg interaction
\cite{Burkard99,Hu00} to anisotropic ones \cite{Meier03}. The
interactions have very short range in the context of these previous
schemes. This often leads to constraints on the architecture
\cite{Lidar02} and serious technical obstacles \cite{Svore05}.
Therefore, it is enlightening to find out one kind of long range
effective interaction for quantum information processing. Recently,
the scheme for a quantum bus of a spin chain has been put forward
\cite{Friesen07}. Through the electrical control of external spin
qubits of Loss and DiVincenzo type \cite{Petta05}, the long range
isotropic Heisenberg interaction is demonstrated in the work
\cite{Friesen07} where the bus is considered as a effective qubit.
However, after each quantum logic operation, the spin bus must be
initialized into its working space. Apparently, this gives rise to
the complexity of the operating mode. In condensed matters, coupled
spin chains are very typical for quantum many body systems.
Different from one single spin-$\frac 12$ chain, two coupled chains
exhibit the natural energy gap \cite{Dagotto96}. The gapped systems
like two coupled spin chains can be experimentally prepared by the
arrays of quantum dots. Here we show how to engineer two coupled
chains as a spin bus for the realization of quantum information
processing. Our method can identify arbitrary single-qubit gates and
controlled-NOT one by electrically controlling the weak coupling
between spin qubits and the bus. During the whole process, the spin
bus cannot share quantum information, which reduces the complication
in the operation sequence.

In this paper, two strongly coupled spin chains with the anisotropic
Heisenberg XXZ exchange interaction are regarded as a bus. The form
of effective interaction is analytically given in Sec. II. The
effective interaction can be tuned by the variations of the
anisotropy and weak coupling between spin qubits and the bus. In
Sec. III, the universal set of quantum logic gates is realized. The
effect of quantum fluctuations in the spin bus is taken into
account. A discussion concludes the paper.

\section{The long range effective interaction}

In our method, the whole quantum system involves a bus of two
coupled chains and external computational spin qubits which are
weakly connected to the bus. For the qubits of the Loss and
DiVincenzo type, the weak coupling can be electrically manipulated.
If the bus system is always frozen at the ground state, the long
range effective interaction can be attained at the computational
qubits where quantum information can be efficiently processed. The
total Hamiltonian in this system is written as
\begin{equation}\label{eq1}
H=H_{0}+H_{in}
\end{equation}
where $H_{0}$ is the Hamiltonian of the bus and $H_{in}$ describes
the weak interaction between computational spin qubits and the bus.
The general case of the bus system $H_{0}$ with the Heisenberg XXZ
anisotropic exchange is considered as
\begin{eqnarray}\label{eq2}
&H_{0}=&J\sum_{i=1}^{2}\sum_{j=1}^{L-1}(s_{i,j}^{x}s_{i,j+1}^{x}+s_{i,j}^{y}s_{i,j+1}^{y}+\Delta_{j}
s_{i,j}^{z}s_{i,j+1}^{z})\nonumber
\\& &+J\sum_{j=1}^{L}\vec{s}_{1,j}\cdot
\vec{s}_{2,j}
\end{eqnarray}
Here $\vec{s}_{i,j}$ denotes the spin operator at the $j$-th site of
the $i$-th chain. The first item of Eq. (2) implies the chain with
the anisotropy $\Delta_{j}=\Delta$ and second item describes the
strong antiferromagnetic coupling between two chains. When the
computational spin qubits $\vec{\tau}_{A}$ and $\vec{\tau}_{B}$ are
weakly coupled to the bus at the connecting nodes
$\vec{s}_{i,j}=\vec{s}_{m}$ and $\vec{s}_{i',j'}=\vec{s}_{n}$, the
interaction can be expressed by $H_{in}=J_{A}\vec{\tau}_{A}\cdot
\vec{s}_{m}+J_{B}\vec{\tau}_{B}\cdot \vec{s}_{n}$.

The bus system has the non-degenerate ground state
$|\psi_{0}\rangle$ with the energy $\epsilon_{0}$ and the first
excited state \cite{Dagotto96}. Due to so large energy gap, the
system can be frozen at the ground state when the physical
temperature $kT$ is much smaller than the gap. The effective
Hamiltonian between two computational qubits $H_{eff}^{(A,B)}$ can
be obtained in this condition. Through the use of the standard
canonical transformation formalism \cite{Ferreira07}, the effective
Hamiltonian of the whole system can be expressed by
\begin{equation}\label{eq3}
H_{eff}=(\langle
\psi_{0}|H|\psi_{0}\rangle+H_{eff}^{(A,B)})\otimes|\psi_{0}\rangle
\langle \psi_{0}|
\end{equation}
Because the total spin at the ground state $|\psi_{0}\rangle$ is
zero, the mean value $\langle
\psi_{0}|H|\psi_{0}\rangle=\epsilon_{0}$. The effective Hamiltonian
$H_{eff}^{(A,B)}=-\sum_{k>0}\sum_{\lambda_{k}=0}^{d_{k}-1}\frac
{\langle
\psi_{0}|H_{in}P_{k}^{\lambda_{k}}H_{in}|\psi_{0}\rangle}{\epsilon_{k}-\epsilon_{0}}$.
The projector $P_{k}^{\lambda_{k}}=|\psi_{k}^{\lambda_{k}}\rangle
\langle \psi_{k}^{\lambda_{k}}|$ where
$|\psi_{k}^{\lambda_{k}}\rangle,(\lambda_{k}=0,1,\cdots,d_{k}-1)$ is
the $k$-th excited degenerate state with the energy $\epsilon_{k}$
and $d_{k}$ is the degree of degeneracy for this state. Thus, the
expression of $H_{eff}^{(A,B)}$ is given by
\begin{equation}\label{eq4}
H_{eff}^{(A,B)}=-\sum_{k,\lambda_{k}}2J_{A}J_{B}\sum_{\alpha}Re(m_{\alpha}n_{\alpha}^{\ast})\tau_{a}^{\alpha}\tau_{b}^{\alpha}
+C_{eff}
\end{equation}
where the constant $C_{eff}=-\sum_{k,\lambda_{k}}
\sum_{\alpha}\frac{J_{A}^{2}}{4}|m_{\alpha}|^{2}+\frac
{J_{B}^{2}}{4}|n_{\alpha}|^{2}$. The parameters $m_{\alpha}=\frac
{\langle
\psi_{0}|s_{m}^{\alpha}|\psi_{k}^{\lambda_{k}}\rangle}{\sqrt{\epsilon_{k}-\epsilon_{0}}}$
and $n_{\beta}=\frac {\langle
\psi_{0}|s_{n}^{\beta}|\psi_{k}^{\lambda_{k}}\rangle}{\sqrt{\epsilon_{k}-\epsilon_{0}}}$.
It is shown that the parameters
$m_{\alpha}(k,\lambda_{k}),n_{\beta}(k,\lambda_{k})$ are determined
by the energy spectra of the bus system. For this transient
invariant quantum system $H_{0}$, the energy spectrum can be
calculated. It is found out that these parameters satisfy the
relation as
\begin{equation}\label{eq5}
-\sum_{k>0}\sum_{\lambda_{k}=0}^{d_{k}-1}m_{\alpha}(k,\lambda_{k})n_{\beta}^{\ast}(k,\lambda_{k})=\gamma_{m,n}^{\alpha}\delta_{\alpha,\beta}
\end{equation}
where $\gamma_{m,n}^{\alpha}$ is real and Eq. (5) is zero if $\alpha
\neq \beta$. Moreover, $\gamma_{m,n}^{x}=\gamma_{m,n}^{y} \neq
\gamma_{m,n}^{z}$ under the assumption of $\Delta \neq 1$. In
general, the effective Hamiltonian of two computational qubits is
written in form of the anisotropic Heisenberg interaction
$H_{eff}^{(A,B)}=2J_{A}J_{B}[\gamma_{m,n}^{x}(\tau_{A}^{x}\tau_{B}^{x}+\tau_{A}^{y}\tau_{B}^{y})+\gamma_{m,n}^{z}\tau_{A}^{z}\tau_{B}^{z}]+C_{eff}$.
It is seen that the effective long range interactions
$J_{m,n}^{\alpha}=2J_{A}J_{B}\gamma_{m,n}^{\alpha}$ primarily rely
on the properties of the bus.

To analytically give the effective interactions, the simplest case
of $L=2$ is considered. The energy spectra and corresponding
eigenstates of $H_{0}$ for $\Delta \in (0,1]$ are spanned in the
product Hilbert space for all of the spins in the bus
\begin{widetext}
\begin{align}\label{eqn6}
\epsilon_{0}&=J\eta_{0}  &|\psi_{0}\rangle&=a_{0}(|\downarrow
\uparrow \downarrow \uparrow \rangle+|\uparrow \downarrow \uparrow
\downarrow \rangle)+b_{0}(|\downarrow \downarrow \uparrow \uparrow
\rangle+|\uparrow \uparrow \downarrow  \downarrow
\rangle)+c_{0}(|\downarrow \uparrow \uparrow \downarrow
\rangle+|\uparrow  \downarrow  \downarrow \uparrow \rangle)\nonumber,\\
 \epsilon_{1}&=-J  &|\psi_{1}^{0}\rangle&=\frac
12(|\downarrow \uparrow \uparrow \uparrow \rangle-| \uparrow
\downarrow \uparrow \uparrow \rangle+| \uparrow \uparrow \downarrow
\uparrow \rangle-| \uparrow \uparrow \uparrow\downarrow \rangle)\nonumber,\\
& &|\psi_{1}^{1}\rangle&=\frac 12(|\uparrow \downarrow \downarrow
\downarrow \rangle-| \downarrow \uparrow \downarrow \downarrow
\rangle+| \downarrow \downarrow \uparrow \downarrow
\rangle-|\downarrow \downarrow
\downarrow \uparrow \rangle)\nonumber,\\
\epsilon_{2}&=-\frac {J}{2}(1+\Delta)  &|\psi_{2}\rangle&=\frac
1{\sqrt{2}}(|\downarrow \uparrow \downarrow \uparrow
\rangle-|\uparrow \downarrow \uparrow \downarrow \rangle)\nonumber,\\
\epsilon_{3}&=-\frac J2(1-\Delta)  &|\psi_{3}\rangle&=\frac
1{\sqrt{2}}(|\downarrow \downarrow \uparrow \uparrow
\rangle-|\uparrow \uparrow \downarrow \downarrow \rangle)\nonumber,\\
\epsilon_{4}&=J\eta_{1}  &|\psi_{4}\rangle&=a_{1}(|\downarrow
\uparrow \downarrow \uparrow \rangle+|\uparrow \downarrow \uparrow
\downarrow \rangle)+b_{1}(|\downarrow \downarrow \uparrow \uparrow
\rangle+|\uparrow \uparrow \downarrow  \downarrow
\rangle)+c_{1}(|\downarrow \uparrow \uparrow \downarrow
\rangle+|\uparrow  \downarrow  \downarrow \uparrow \rangle)\nonumber,\\
\epsilon_{5}&=0 &|\psi_{5}^{0}\rangle&=\frac 12(|\downarrow \uparrow
\uparrow \uparrow \rangle+| \uparrow \downarrow \uparrow \uparrow
\rangle-| \uparrow \uparrow \downarrow
\uparrow \rangle-| \uparrow \uparrow \uparrow\downarrow \rangle)\nonumber,\\
 & &|\psi_{5}^{1}\rangle&=\frac 12(|\downarrow \uparrow
\uparrow \uparrow \rangle-| \uparrow \downarrow \uparrow \uparrow
\rangle-| \uparrow \uparrow \downarrow
\uparrow \rangle+| \uparrow \uparrow \uparrow\downarrow \rangle)\nonumber,\\
 & &|\psi_{5}^{2}\rangle&=\frac 12(|\uparrow \downarrow \downarrow
\downarrow \rangle+| \downarrow \uparrow \downarrow \downarrow
\rangle-| \downarrow \downarrow \uparrow \downarrow
\rangle-|\downarrow \downarrow
\downarrow \uparrow \rangle)\nonumber,\\
 & &|\psi_{5}^{3}\rangle&=\frac 12(|\uparrow \downarrow \downarrow
\downarrow \rangle-| \downarrow \uparrow \downarrow \downarrow
\rangle-| \downarrow \downarrow \uparrow \downarrow
\rangle+|\downarrow \downarrow
\downarrow \uparrow \rangle)\nonumber,\\
\epsilon_{6}&=\frac J2(1-\Delta) &|\psi_{6}\rangle&=\frac
1{\sqrt{2}}(|\downarrow \uparrow \uparrow \downarrow
\rangle-|\uparrow \downarrow \downarrow \uparrow  \rangle)\nonumber,\\
\epsilon_{7}&=\frac J2(1+\Delta) &|\psi_{7}^{0}\rangle&=|\downarrow
\downarrow \downarrow \downarrow \rangle\nonumber,\\
 & &|\psi_{7}^{1}\rangle&=|\uparrow \uparrow \uparrow \uparrow
\rangle\nonumber,\\
\epsilon_{8}&=J &|\psi_{8}^{0}\rangle&=\frac 12(|\downarrow \uparrow
\uparrow \uparrow \rangle+| \uparrow \downarrow \uparrow \uparrow
\rangle+| \uparrow \uparrow \downarrow
\uparrow \rangle+| \uparrow \uparrow \uparrow\downarrow \rangle)\nonumber,\\
& &|\psi_{8}^{1}\rangle&=\frac 12(|\uparrow \downarrow \downarrow
\downarrow \rangle+| \downarrow \uparrow \downarrow \downarrow
\rangle+| \downarrow \downarrow \uparrow \downarrow
\rangle+|\downarrow \downarrow
\downarrow \uparrow \rangle)\nonumber,\\
\epsilon_{9}&=J\eta_{2}  &|\psi_{9}\rangle&=a_{2}(|\downarrow
\uparrow \downarrow \uparrow \rangle+|\uparrow \downarrow \uparrow
\downarrow \rangle)+b_{2}(|\downarrow \downarrow \uparrow \uparrow
\rangle+|\uparrow \uparrow \downarrow  \downarrow
\rangle)+c_{2}(|\downarrow \uparrow \uparrow \downarrow
\rangle+|\uparrow  \downarrow  \downarrow \uparrow \rangle)
\end{align}
\end{widetext}
Here the basis of the space $|\uparrow \downarrow \uparrow
\downarrow \rangle$ labels
$|\uparrow_{1,1}\rangle|\downarrow_{1,2}\rangle|\uparrow_{2,2}\rangle|\downarrow_{2,1}\rangle$
where $|\uparrow(\downarrow)_{i,j}\rangle$ is the eigenstate of the
spin operator $s_{i,j}^{z}$ with the corresponding eigenvalues $\pm
\frac 12$. The coefficients of Eq. (6) are given by $a_{f}=\frac
{1}{\sqrt{2[1+1/(\eta_{f}+\frac {1-\Delta}{2})^{2}+1/(\eta_{f}-\frac
{1-\Delta}{2})^{2}]}}, b_{f}=a_{f}/(\eta_{f}+\frac
{1-\Delta}{2}),c_{f}=a_{f}/(\eta_{f}-\frac {1-\Delta}{2})$ and $\{
\eta_{f=0,1,2}|\eta_{0}<\eta_{1}<\eta_{2}\}$ are the three real
roots for the equation of $\eta^{3}+\frac
{1+\Delta}{2}\eta^{2}-[2+\frac {(1-\Delta)^2}{4}]\eta-\frac
{(1-\Delta)^{2}(1+\Delta)}{8}=0$. For the special case of
$\Delta=1$, the roots $\{\eta_{f=0,1,2} \}=\{-2,0,1\}$ and the
corresponding coefficients are obtained as $a_{0}=\frac
1{\sqrt{3}},b_{0}=c_{0}=-\frac {1}{2\sqrt{3}}$,
$a_{1}=0,b_{1}=c_{1}=-\frac {1}{2}$ and $a_{2}=b_{2}=c_{2}=\frac
1{\sqrt{6}}$. Therefore, if the computational qubits are weakly
coupled to the bus, we can calculate the effective interactions by
$\gamma_{(1,1),(2,1)}^{\alpha}=\gamma_{1,2}^{\alpha}=\frac {1}{6J}$,
$\gamma_{(1,1),(2,2)}^{\alpha}=\gamma_{1,3}^{\alpha}=-\frac {1}{8J}$
and $\gamma_{(1,1),(1,2)}^{\alpha}=\gamma_{1,4}^{\alpha}=\frac
{1}{6J}$. In the condition of $\gamma_{m,n}^{\alpha}<0$, the
antiferromagnetic interactions
$J_{m,n}^{\alpha}=2J_{A}J_{B}\gamma_{m,n}^{\alpha}$ can be attained
if the weak couplings are chosen as $J_{A}J_{B}<0$. The effects of
the anisotropy on the long range interactions are shown by Fig. 1.
It is seen that the effective interaction $\gamma_{1,2}^{x}$ is
almost linearly decreased with the increase of the anisotropy
$\Delta$ in Fig. 1(a). Here the effective anisotropy
$\Delta_{1,2}=\gamma_{1,2}^{z}/\gamma_{1,2}^{x}$ is also illustrated
by Fig. 1(b). These values almost take on the exponential increase.
At the point of $\Delta=1$, the effective anisotropy is the maximal
value of one where the isotropic Heisenberg interaction occurs. The
connecting nodes of the bus are labeled in the new sequence given by
Fig. 2(a). Similar to the results of \cite{Friesen07}, the effective
interactions $J_{1,n}^{\alpha}$ alternately occurs between the
antiferromagnetic and ferromagnetic ones. To clearly show this
character, the effective interaction is plotted with respect to the
distance $n-1$ between two nodes by Fig. 2. It is found that the
coefficient of the effective interaction $\gamma_{1,n}^{x}>0$ when
the distance $n-1$ is odd. And the values of $\gamma_{1,n}^{x}$
decays almost exponentially. The opposite case is shown in the
condition that $n-1$ is even. With the increase of the length of the
chains, we also calculate the effective interactions. From Fig.
2(c), it is seen that $J_{1,2}^{x}$ is gradually increased to the
certain value with length $L$. This fact demonstrates that the
effective interaction can also be obtained in coupled chains with
infinite size.

\section{The realization of quantum logic operations}

After the evolution of the system, quantum information will be
manipulated in the computational qubits under the assumption that
the bus is always frozen at the ground state. It is the advantage
that the distances of quantum logic operations can be enlarged much
more. To implement the universal quantum computation, we need
realize a set of quantum logic gates which consist of arbitrary
single-qubit and controlled-NOT one. According to \cite{Meier03}, an
efficient way to construct the exact controlled-NOT logic gate can
be provided for the general Heisenberg XXZ interaction
\begin{equation}\label{eq7}
U_{CNOT}^{(A,B)}=R_{B}^{y}(\frac {\pi}{2})U_{CPF}R_{B}^{y}(\frac
{\pi}{2})^{\dag}
\end{equation}
where the controlled phase flip gate $U_{CPF}=\exp[i\frac
{2\pi}{3}\vec{n_{1}}\cdot (\vec{\tau}_{A}+\vec{\tau}_{B})/\hbar]
\exp[i\frac {2\pi}{3}\vec{n_{2}}\cdot
(\vec{\tau}_{A}+\vec{\tau}_{B})/\hbar]U(\frac
{\pi}{2})R_{A}^{y}(-\pi) U(\frac {\pi}{2})R_{A}^{x}(\frac
{\pi}{2})R_{B}^{x}(\frac {\pi}{2})$. Here
$R_{i}^{\alpha}(\theta)=\exp(-i\theta \tau_{i}^{\alpha}/\hbar)$ is
the single-qubit rotation at the computational qubit $i=A(B)$. The
unit directions are given by $\vec{n}_{1}=\frac
{1}{\sqrt{3}}(\vec{e}_{x}-\vec{e}_{y}+\vec{e}_{z}),
\vec{n}_{2}=\frac
{1}{\sqrt{3}}(\vec{e}_{x}+\vec{e}_{y}-\vec{e}_{z})$. And the unitary
operation $U(\frac {\pi}{2})=\exp[\frac
{-i}{\hbar}\int_{0}^{t_{c}}H_{eff}^{(A,B)}dt]$ where $\frac
{\pi}{2}=\int_{0}^{t_{c}}J_{m,n}^{x}dt/\hbar$. For the case of
$\Delta=1$, the exact swap gate is obtained by $U_{swap}=\exp[\frac
{-i}{\hbar}\int_{0}^{t_{s}}H_{eff}^{(A,B)}dt]$ and the swap time
$t_{s}=\pi \hbar/J_{m,n}^{\alpha}$. From Eq. (7), it is seen that
the controlled-NOT gate needs to be implemented by means of some
single-qubit operations.

It is necessary to study how to operate the single-qubit rotations.
When only one computational qubit $\vec{\tau}_{A}$ is coupled to the
node $\vec{s}_{m}$ of the bus and the local magnetic field $\vec{b}$
is applied, the hamiltonian of the whole system is written by
$H=H_{0}+\vec{b}\cdot \vec{\tau}_{A}$. Similarly, the effective
hamiltonian can also be given by
\begin{equation}\label{eq8}
H_{eff}=(\epsilon_{0}+\vec{b}\cdot \vec{\tau}_{A}+\frac
{J_{A}}{4}\sum_{\alpha=x,y,z}\gamma_{m,m}^{\alpha})\otimes
|\psi_{0}\rangle \langle \psi_{0}|
\end{equation}
If the local magnetic field
$\vec{b}=b_{\alpha}\vec{e}_{\alpha}(\alpha=x,y,z)$ is chosen, the
unitary operation after the certain time is expressed by
\begin{equation}\label{eq9}
u(\theta)=\exp(-\frac {i}{\hbar}\int
H_{eff}dt)=R_{A}^{\alpha}(\theta)\cdot \exp(-\frac {i}{\hbar}\Phi)
\end{equation}
where $\theta=-\int b_{\alpha}dt,\Phi=\int (\epsilon_{0}+\frac
{J_{A}}{4}\sum_{\alpha=x,y,z}\gamma_{m,m}^{\alpha})dt$. It is shown
that the single-qubit unitary rotation is operated by one
computational qubit in the local field which is connected to the
bus. Therefore, the universal quantum computation can be realized in
the bus of two coupled chains.

It is noted that during the whole process of quantum logic
operations the adiabatic criterion is satisfied in order to avoid
the crossing of the energy levels. According to \cite{Friesen07} ,
the adiabatic criterion can be expressed by
$t_{s}(\epsilon_{1}-\epsilon_{0})>2\pi\hbar$ for the case of
$\Delta=1$. Owing to the previous work of \cite{White94}, the energy
gap of the bus $\epsilon_{1}-\epsilon_{0}\sim \frac
{J}{2}+CJ\exp(-\frac {L}{4})/L$. From the above calculation,
$t_{s}(\epsilon_{1}-\epsilon_{0})>\frac
{J\pi\hbar}{4J_{A}J_{B}\gamma_{m,n}^{\alpha}}$. When the parameters
$J_{A}=J_{B}=J/10$, the value of
$J/4J_{A}J_{B}\gamma_{m,n}^{\alpha}$ is about $10^{2}$ for coupled
infinite chains. That is, the adiabatic criterion can be achieved
$t_{s}(\epsilon_{1}-\epsilon_{0})>100\pi\hbar>2\pi\hbar$. This shows
that coupled chains with infinite size can be served as the bus for
quantum information processing. If the actual physical parameters
are considered as $J\simeq 1 \quad mev, J_{A}=J_{B}=1/100,\Delta=1$,
the operation time for two nodes $m=(1,1),n=(2,1)$ is $t_{c}\simeq
27\quad ps$.

In reality, the certain quantum fluctuations always exist in the
bus. To clearly show the effects of the fluctuations on quantum
computation, small variations of the anisotropy in the nodes of
$\vec{s}_{m},\vec{s}_{n}$ are considered by
$\Delta_{m}=\Delta(1+\delta_{m})$ and
$\Delta_{n}=\Delta(1+\delta_{n})$. It is found that the effective
long range interaction can also be written in the form of Heisenberg
XXZ exchange under this assumption. The norm of the operator
$n=||U(\delta_m,\delta_n)-U(0,0)||$ is used to evaluate the error of
the operators where $U(\delta_m,\delta_n)$ is the evolution operator
of $H_{eff}^{(A,B)}$ after the time
$t_{c}=\pi\hbar/4J_{A}J_{B}\gamma_{m,n}^{x}$ in the condition of the
quantum fluctuations. And $U(0,0)$ corresponds to the one without
the fluctuations. The error can be analytically given by
\begin{equation}\label{eq10}
n=\max \{\sqrt{2[1-\cos\frac {\pi\delta_x}{4}]}, \sqrt{2[1-\cos\frac
{\pi\delta_z}{4}]} \}
\end{equation}
where the parameters $\delta_x=\frac
{\gamma_{m,n}^{x}(0,0)-\gamma_{m,n}^{x}(\delta_m,\delta_n)}{\gamma_{m,n}^{x}(0,0)}$
and $\delta_z=\frac
{\gamma_{m,n}^{z}(0,0)-\gamma_{m,n}^{z}(\delta_m,\delta_n)}{\gamma_{m,n}^{x}(0,0)}$.
Here $\gamma_{m,n}^{\alpha}(\delta_m,\delta_n)$ is the coefficient
of the effective interaction with the fluctuations of the anisotropy
while $\gamma_{m,n}^{\alpha}(0,0)$ is the one without the
fluctuations. The error of the operator is shown in Fig. 3. It is
found that the error arrives at the minimum when
$\delta_m=-\delta_n$. For a certain value of $\delta_m$, the error
is increased with $\delta_n$. In the range of
$|\delta_m|,|\delta_n|<0.005$, the order of the error can be
decreased to $10^{-5}$ which is well below than the threshold
$10^{-4}$ for the fault-tolerant quantum computation.

\section{Discussion}

The gapped spin system of two coupled chains is served as the bus.
During the whole process, the bus can always be frozen at the ground
state for the large energy gap. By weakly connecting external
computational spin qubits with the bus, the effective long range
interactions can be attained. If two computational qubits are
coupled, the effective hamiltonian has the form of anisotropic
Heisenberg XXZ which can be useful for the construction of the
controlled-NOT gate. It is found that the effective interaction
alternately occurs between the antiferromagnetic and ferromagnetic
ones. The effective interaction can be constructed in coupled chains
with infinite size. The values decay almost exponential with
increasing the distance of two nodes. In regard to only one qubit
connected to the bus, single-qubit rotations can be implemented by
the local magnetic field. Therefore, the universal quantum
computation can be realized in the long range which can eliminate
the technical difficulty arising from the exchange of short
distances. Meanwhile, compared with the proposal of
\cite{Friesen07}, the operating sequence of quantum information
processing can be simplified. The quantum logic gates with small
errors can be achieved in the condition of the quantum fluctuations
of the anisotropy of coupled chains.

\section{Acknowledgement}

It is a pleasure to thank Yinsheng Ling, Jianxing Fang and Qing
Jiang for their many fruitful discussions about the topic. The
financial support from the special Research Fund for the Doctoral
Program of Higher Education(Grant No. 20050285002) and the National
Science Fund of China (Grand No. 10774108) are gratefully
acknowledged.

\newpage

{\Large \bf Figure Captions}

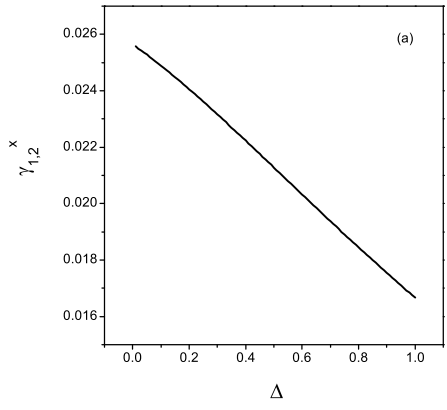
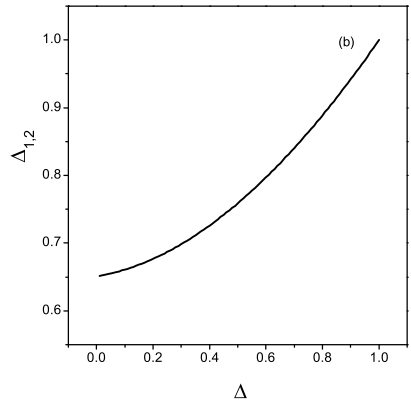
{\bf Fig. 1}

The effective interactions of $L=2$  are numerically plotted as a
function of the anisotropy when the coupling $J=10$, $J_{A}=J_{B}=1$
(a). The coefficient of the effective interaction $\gamma_{1,2}^{x}$
is shown; (b). The effective anisotropy $\Delta_{1,2}$ is also
illustrated.

{\bf Fig. 2}

(a). The new sequence of the connecting nodes are labeled by the
solid lines of the bus where the dash lines denote the couplings;
(b). When one computational qubit is connected to the node $m=1$,
the coefficient of the effective interaction $\gamma_{1,n}^{x}$ is
plotted as the distance $n-1$ for $J=10$, $J_{A}=J_{B}=1$,
$\Delta=0.2$ and $L=6$; (c). The effective interaction $J_{1,2}^{x}$
is plotted with increasing the length $L$ of chains for $J=10$,
$J_{A}=J_{B}=1$, $\Delta=0.2$.

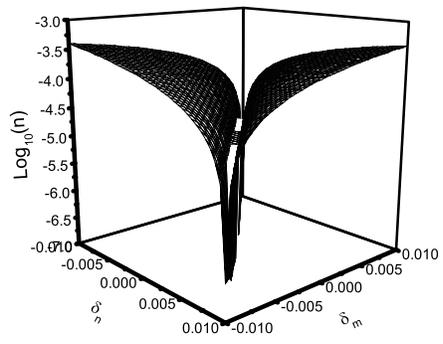
{\bf Fig. 3}

The logarithm of the error $\log_{10}(n)$ in the chains with $L=4$
is plotted as a function of the anisotropy for $J=10$,
$J_{A}=J_{B}=1$, $\Delta=0.2$


\begin{references}
\bibitem{Divin98} D. Loss and D. P. DiVincenzo, Phys. Rev. A
\textbf{57}, 120(1998).
\bibitem{Kane98} B. E. Kane, Nature(London), \textbf{393},
133(1998).
\bibitem{Hu00} X. Hu and S. Das Sarma, Phys. Rev. A \textbf{61},
062301(2000).
\bibitem{Burkard99} G. Burkard, D. Loss and D. P.
DiVincenzo, Phys. Rev. B \textbf{59}, 2070(1999).
\bibitem{Meier03} F. Meier, J. Levy and D. Loss, Phys. Rev. Lett.
\textbf{90}, 047901(2003).
\bibitem{Lidar02} D. A. Lidar and R. M. Noack, Phys. Rev. Lett.
\textbf{88}, 017905(2002).
\bibitem{Svore05} K. M. Svore, B. M. Terhal and D. P. DiVincenzo, Phys.
Rev. A \textbf{72}, 022317(2005).
\bibitem{Friesen07} M. Friesen, A. Biswas, X. Hu, and D. Lidar,
Phys. Rev. Lett. \textbf{98}, 230503(2007).
\bibitem{Petta05} J. R. Petta, A. C. Johnson, J. M. Taylor, E. A. Laird, A. Yacoby, M. D. Lukin, C. M. Marcus, M. P. Hanson and A. C.
Gossard, Science \textbf{309}, 2180(2005).
\bibitem{Dagotto96} E.
Dagotto and T. M. Rice, Science \textbf{271}, 618(1996).
\bibitem{Ferreira07} A. Ferreira, J. M. B. Lopes dos Santos, arxiv:
quant-ph/0708.0320v1(2007).
\bibitem{White94} S. R. White, Phys. Rev. Lett.
\textbf{73}, 886(1994).
\end{references}
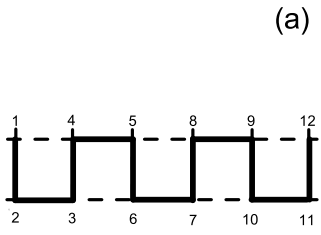
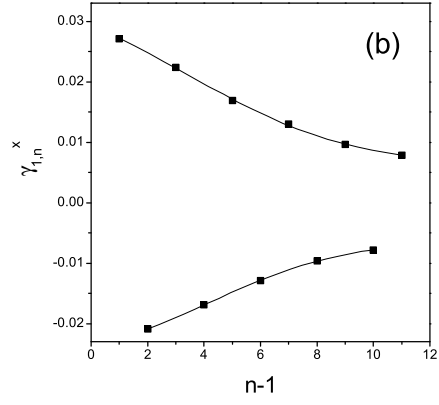
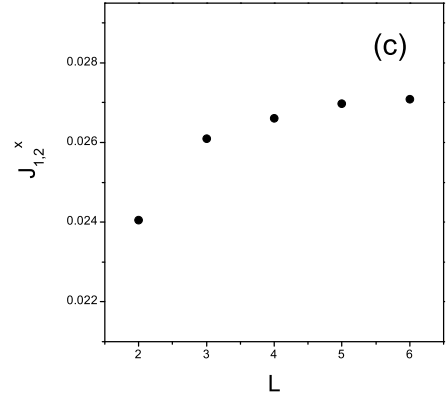
\end{document}